\documentstyle[aps]{revtex}
\begin{document}

\thispagestyle{empty}
\title{ Complex sine-Gordon Equation  \\
in Coherent Optical Pulse Propagation  }
\author{Q-Han Park\footnote{ E-mail address; qpark@nms.kyunghee.ac.kr }
{and}
H. J. Shin\footnote{ E-mail address; hjshin@nms.kyunghee.ac.kr }}
\address{\begin{center}{\it  
Department of Physics \\
and \\
Research Institute of Basic Sciences \\
Kyunghee University, 
Seoul, 130-701, Korea}\end{center}}
\maketitle
\begin{abstract}
It is shown that the McCall-Hahn theory of self-induced  transparency 
in coherent optical pulse propagation can be identified with the complex 
sine-Gordon theory in the sharp line limit. We reformulate the theory in 
terms of the deformed gauged Wess-Zumino-Witten sigma model and address 
various new aspects of self-induced transparency.
\bigskip
\end{abstract}
\maketitle
Self-induced transparency(SIT), a phenomenon of anomalously low 
energy loss in coherent optical pulse propagation, was first discovered 
by McCall and Hahn\cite{McCall} and  the integrability of the SIT equation 
 was demonstrated by employing the inverse scattering method\cite{AKN}. 
When phase variation is ignored in the case for a symmetric frequency 
distribution $g(\Delta w)$ of inhomogeneous broadening,  McCall and Hahn have 
proved an area theorem for pulse propagation. In the sharp line limit where 
the pulse spectral width is much broader than the atomic line width such that 
$g(\Delta w) = \delta (w-w_{0})$, the SIT equation reduces to the well-known sine-Gordon equation and the $2\pi $ area pulse becomes 
a 1-soliton  of the sine-Gordon theory. However, when phase variation 
is included, the area theorem no longer holds and the structure of 
SIT in general has not been well understood except for the 
construction of explicit solutions by the inverse scattering method\cite{AKN}\cite{lamb}. 
In view of various physical situations where phase variations become important, e.g. 
chirping effect or the slow reshaping of pulses in the off resonant 
case\cite{diels}, it is important to undestand the analytic properties of SIT 
with phase variation. In addition, despite 
its integrability, the SIT theory in the sharp line limit has not been 
identified with a known 1+1 dimensional integrable field theory, which made 
 a systematic understanding of SIT in terms of a lagrangian 
field theory impossible.

The purpose of this Letter is to show that the SIT theory 
with phase variation can be 
identified with the complex sine-Gordon theory in the sharp line 
limit. The complex sine-Gordon theory, a generalization 
of the sine-Gordon theory 
with a phase degree of freedom, can be reformulated in terms of a 
nonlinear sigma model which is known as the integrably deformed gauged 
Wess-Zumino-Witten(WZW) model associated with the coset 
SU(2)/U(1)\cite{park}\cite{shin1}. This allows us to address 
various new aspects of SIT 
in terms of characteristics of the complex sine-Gordon theory; e.g. 
topological v.s. non-topological solitons, local gauge symmetry,  
the U(1)-charge conservation, the chiral symmetry and the Krammers-Wannier 
duality for dark v.s. bright solitons. In physical terms, the local 
gauge transformation accounts for the off-resonance effect of SIT as well 
as the inhomogeneous broadening of medium. The U(1)-charge characterizes 
the effect of phase variation in optical pulses. We also address how our 
effective potetial formulation can be extended to more physical cases 
without the sharp line limit.

The SIT equation is given by
\begin{eqnarray} 
\bar{\partial} E + 2 \beta <P> &=& 0 \nonumber \\
\partial D - E^{*}P - EP^{*} &=& 0 \nonumber \\
\partial P + 2i\Delta w P + 2ED &=& 0
\end{eqnarray} 
where $\Delta w = w-w_{0} \ , \ \partial \equiv \partial /\partial z \ , \ \bar{\partial} \equiv \partial / \partial \bar{z} \ ,
z= t-x/c , \bar{z} = x/c$.  $E,P$ and $ D$ represent the electric field, 
the polarization and the population inversion respectively. The bracket 
denotes an averaging over the distribution function of inhomogeneous 
broadening. Since  Eq.(1) is invariant under the interchange 
$(\beta , E, P , D) \leftrightarrow (-\beta ,E, -P , -D) $, 
we assume the coupling 
constant $\beta $ to be positive which we set to one by rescaling $E, P$ and 
$ D$. In a simpler case where phase variation is ignored to make $ E$ real 
and the frequency distribution is sharply peaked at 
the carrier frequency($\Delta w =0$), we may parametrize $E, P$ and $ D$ by
\begin{equation}
E=E^{*} = \partial \varphi \ \ , \  \ <P> = P = -\sin{2\varphi } \ \ , \ \ 
D= \cos{2\varphi } \ .
\end{equation}
Then the SIT equation reduces to the well-known 
sine-Gordon equation
\begin{equation}
\bar{\partial}\partial \varphi - 2\beta \sin{2\varphi } = 0 .
\end{equation}
In order to include phase variation as well as the off-resonance 
effect ($\Delta w \ne 0$), we assume $E$ to be complex and require that the 
distribution is sharply peaked not necessarily at the carrier frequency, i.e.  
$g(\Delta w) = \delta (\Delta w - \xi )$ for some constant $\xi$.  Introduce a 
more general parameterization 
of $E, P$ and $D$ in terms of three scalar fields $\varphi , 
\theta $ and $\eta $,
\begin{equation}
E = e^{i(\theta - 2\eta )}( 2\partial \eta {\cos{\varphi } 
\over \sin{\varphi }} - i\partial \varphi ) 
\ \ , \ \ 
P = ie^{i(\theta - 2\eta )}\sin{2\varphi } 
\ \ , \ \ 
D = \cos{ 2\varphi } \ .
\end{equation}
The main result is that the SIT equation given in Eq.(1) then changes into a 
couple of second order nonlinear differential equations known as the complex 
sine-Gordon equation,
\begin{eqnarray} 
\bar{\partial}\partial \varphi + 4{\cos{\varphi } \over 
\sin^{3}{\varphi }}\partial \eta \bar{\partial} 
\eta -2\sin{2\varphi } & =& 0 
\\
\bar{\partial} \partial \eta - {2 \over \sin{2\varphi }}
(\bar{\partial} \eta \partial \varphi + \partial \eta \bar
{\partial} \varphi ) &=& 0
\end{eqnarray} 
together with a couple of first order constraint equations,
\begin{eqnarray} 
2\cos^{2}{\varphi }\partial \eta - \sin^{2}{\varphi }\partial 
\theta - 2\xi \sin^{2}{\varphi } 
&= & 0 \\
2\cos^{2}{\varphi }\bar{\partial} \eta + \sin^{2}{\varphi 
}\bar{\partial} \theta  &=& 0 \ .
\end{eqnarray} 

The complex sine-Gordon equation first appeared in 1976 in a description 
of relativistic vortices in a superfluid \cite{lund}, and also 
independently in a treatment of O(4) nonlinear sigma model\cite{pohl}. 
Note that Eqs.(5) and (6) consistently reduce to the sine-Gordon equation 
when phase variation is ignored so that $\eta =0, \ \theta = {\pi \over 2}$ 
and  the system is on resonance($\xi = 0$). Earlier works on the complex 
sine-Gordon theory have focused only on Eqs.(5) and (6). However, the 
constraints in Eqs.(7) and (8) which are new expressions constitute an 
essential part of the SIT theory, particularly in connection with a local 
U(1)-gauge symmetry of SIT as explained later. Thus we will call 
Eqs.(5)-(8) as the complex sine-Gordon theory. The gauge symmetry structure 
as well as the integrability of Eqs.(5)-(8) may be best understood if we 
reformulate the complex sine-Gordon theory in terms of the action principle. 
The proper action is given by a group theoretical nonlinear sigma model 
action known as the deformed gauged WZW action defined as follows; 
\begin{eqnarray} 
S(g,A, \bar{A} , \beta ) &=& S_{WZW}(g) +
{1 \over 2\pi }\int \mbox {Tr} (- A\bar{\partial} g g^{-1} + 
\bar{A} g^{-1} \partial g
 + Ag\bar{A} g^{-1} - A\bar{A} ) -S_{\mbox{potential}} \nonumber \\
S_{\mbox{potential}}&=& {\beta \over 2\pi }\int \mbox{Tr}gT g^{-1} \bar{T}
\end{eqnarray} 
where $S_{WZW}(g)$ is the conventional SU(2)-WZW action and $g $ is an 
SU(2) matrix function. Tr denotes the trace and $ T = -\bar{T} = i\sigma_{3} = 
\mbox{diag}(i, -i)$ for Pauli matrices $\sigma_{i} $. The local gauge fields 
$A = a(z, \bar{z} )\sigma_{3}, \bar{A} = \bar{a} (z,\bar{z} )\sigma_{3} $  
are introduced to 
gauge the U(1) subgroup of SU(2). Owing to the absence of the kinetic terms, 
the gauge fields $A, \bar{A} $ act as Lagrange multipliers which result in 
the constraint equations. One of the nice properties of our fomulation is that 
the equation of motion arising from the action (9) takes a zero 
curvature form,
\begin{equation}
\delta _{g}S = -{1 \over 2\pi }\int \mbox{Tr}
[\  \partial + g^{-1} \partial g + g^{-1} A g + \beta \lambda T \ , \ \bar{\partial} + \bar{A} + 
{1 \over \lambda }g^{-1} \bar{T} g \ ]g^{-1} \delta g  = 0
\end{equation}
where the constant $\lambda $ is a spectral parameter and the 
square bracket denotes the 
commutation. The constraint equations coming from the 
$A, \bar{A} $-variations are
\begin{eqnarray} 
\delta _{A}S &=& {1 \over 2\pi }\int \mbox{Tr} ( \ - \bar{\partial} g 
g^{-1} + g\bar{A} g^{-1} - \bar{A} \  )
\delta A= 0  \\
\delta _{\bar{A} }S &=& {1 \over 2\pi }\int \mbox{Tr} ( \  g^{-1} 
\partial g  +g^{-1} A g - A \ )\delta \bar{A} = 0 \ .
\end{eqnarray} 
The action (9) is known to possess the local U(1)-vector gauge symmetry 
under the transform;
\begin{equation}
g \rightarrow h^{-1}gh \ \ , \ \ A \rightarrow A + h^{-1}\partial h \ \ , \ \ 
 \bar{A} \rightarrow \bar{A}  + h^{-1}\bar{\partial} h 
\end{equation}
where $h = \exp (\phi (z, \bar{z}) \sigma_{3})$, as well 
as the global U(1)-axial vector 
gauge symmetry under $g \rightarrow hgh $ for a constant $h$. 
In order to identify Eqs.(10)-(12) with the SIT equation, we fix the vector 
gauge by choosing 
\begin{equation}
A = \xi T \ , \ \bar{A} =0
\end{equation}
for a constant $\xi $. Such a gauge fixing is possible due to the flatness of 
$A, \bar{A}$\cite{shin1}. Also, we parameterize the $2\times 2$ matrix $g$ by
\begin{equation}
g=e^{i\eta \sigma_{3}}e^{i\varphi (\cos{\theta }\sigma_{1} 
-\sin{\theta }\sigma_{2})}e^{i\eta 
\sigma_{3}}= \pmatrix{ e^{2i\eta }\cos{\varphi } & 
i\sin{\varphi }e^{i\theta } \cr
i\sin{\varphi }e^{-i\theta } & e^{-2i\eta }\cos{\varphi } }\ .
\end{equation}
Then the parametrization in Eq.(4) arises from an identification of 
$E, P$ and $D$ with $g$ through the relation
\begin{equation}
g^{-1}\partial g + \xi g^{-1}Tg - \xi T = \pmatrix{ 0 & -E  \cr 
E^{*} & 0 } \ \ , \ \ 
g^{-1}\bar{T}g = -i \pmatrix{ D & P \cr P^{*} & -D }
\end{equation}
where we have used the constraint equation (12). Also, the zero 
curvature equation (10) with the identification in Eq.(16) becomes
\begin{equation}
\left[ \  \partial + \pmatrix{ i\beta \lambda + i\xi & -E \cr E^{*} & 
-i\beta \lambda -i\xi } \ , \ 
\bar{\partial} - {i \over \lambda } \pmatrix{D & P \cr P^{*} & -D }
\right]  = 0
\end{equation}
whose components agree precisely with the SIT equation in the 
sharp line limit. The constraint 
equations (11) and (12), combined with Eq.(14), also reduce to Eqs.(7) 
and (8). Thus we have shown that the SIT equation consistently arises from the 
action (9) with the gauge fixing in Eq.(14). Moreover, the zero curvature 
equation (17) demonstrates the integrability of the SIT equation.
The potential term in Eq.(9) changes into the population inversion $D$,
\begin{equation}
S_{\mbox{potential}} = \int {\beta \over \pi }\cos{2\varphi } = 
\int {\beta \over \pi }D   ,
\end{equation}
which for $\beta =1 $ possesses degenerate vacuua at
\begin{equation}
\varphi = \varphi_{n} = (n+ {1\over 2} )\pi , \ n \in Z  \ \mbox{ and } \  
\theta \ = \theta _{0} \ \ \mbox{for} \ \ \theta _{0} \  \mbox{constant} \ .
\end{equation}
The soliton solutions interpolating different vacuua can be obtained 
either by applying the dressing method or by using the B\"{a}cklund 
transformation\cite{shin1}.
In particular, the 1-soliton solution is given by
\begin{eqnarray} 
\cos{\varphi } &=& {b \over \sqrt{(a-\xi )^{2} + b^{2} }} 
\mbox{sech} (2bz -2b C \bar{z} ) \nonumber \\
\eta &=& (a-\xi )z + (a-\xi ) C\bar{z} \nonumber \\
\theta &=& - \tan^{-1}[ {a-\xi \over b } \mbox{coth} 
(2bz - 2bC\bar{z} )] - 2\xi z + 2D\bar{z} 
\end{eqnarray} 
where $a, b$ are arbitrary constants and 
\begin{equation}
C = {1 \over (a-\xi )^{2} + b^{2}} \ , \ D = 0 .
\end{equation}
In terms of $E$,
\begin{equation}
E = -2ib \  \mbox{sech} (2bz - 2bC \bar{z} ) 
e^{-2i(az -D\bar{z}  + (a-\xi )C\bar{z} )} .
\end{equation}
If $a-\xi =0$, this solution interpolates between two different vacuua 
$\varphi_{n}$ and $\varphi_{n+1}$, i.e. it becomes a topological 1-soliton 
($\Delta n =1$).  On resonance where $a = \xi =0$, Eq.(20) reduces to 
the 1-soliton of the sine-Gordon equation, or a $2\pi $ pulse of SIT. 
If $a -\xi \ne 0$, the solution in Eq.(20) reaches to the same vacuum 
asymptotically as $x \rightarrow \pm \infty $ 
so that the topological number is zero ($\Delta  n=0$). 
Nevertheless, except for the topological number, this solution possesses 
all the properties of a soliton so that we call it a nontopological 
1-soliton. It represents a localized pulse with a steadily varying 
phase. The time area of the pulse, which is defined by
\begin{equation}
\Delta S = 2\int |E| dt ,
\end{equation}
is still $2\pi $. It is important to mention that this $2\pi $ area is 
a mere coincidence and should not be confused with the $2\pi $ area of 
the topological one. Because of the interference between phases of each 
nontopological solitons, multi-nontopological solitons in general 
do not possess the area which is an integer multiple of $2\pi $ 
and the area theorem of McCall and Hahn in the case of inhomogenous 
broadening does not hold. 
The stability of nontopological solitons, unlike the topological case whose 
stability is due to the topological protection, arises from the U(1)-charge 
conservation law. Recall that the action (9), consequently Eqs.(5)-(8), 
are invariant under the axial vector transform $g \rightarrow hgh$ or, 
equivalently,
\begin{equation}
\eta \rightarrow \eta + \epsilon \ \   \mbox{ for } 
\epsilon  \  \mbox{constant} .
\end{equation}
The corresponding Noether currents and the charge are given by
\begin{eqnarray} 
J &=& {\cos^{2}{\varphi } \over \sin^{2}{\varphi }}\partial \eta  \ \ , \ \ 
\bar{J} = {\cos^{2}{\varphi } \over \sin^{2}{\varphi }}\bar{\partial} \eta \nonumber \\
Q &=& \int_{-\infty }^{\infty }dx (J + \bar{J}) 
\end{eqnarray} 
where $J, \bar{J}$ satisfy the conservation law,
 $  \partial \bar{J} + \bar{\partial} J = 0$.
In particular, the charge of the 1-soliton in Eq.(20) is
\begin{equation}
Q_{\mbox{1-sol}} = -c(\mbox{sign}[b\cdot (a-\xi )]{\pi \over 2 } - \tan^{-1}{a-\xi \over b}) .
\end{equation}
The stability of nontopological solitons can be shown either by using 
conservation laws in terms of charge and energy as given in \cite{shin1}, 
or by studying the behavior against small fluctuations\cite{shin3}. 

The action (9) also possesses two different types of discrete 
symmetries. This can be seen most easily in a different gauge where 
$A=\bar{A} =0$ which is connected to the gauge in Eq.(14) by an appropriate  
vector gauge transformation.
The first case is {\it the chiral symmetry} under the interchange,
\begin{equation}
z \leftrightarrow \bar{z} \  \ \mbox{and} \ \ 
 \ g \leftrightarrow g^{-1} ( \ \mbox{ or } \eta \leftrightarrow -\eta 
\ , \ \varphi \leftrightarrow -\varphi )
\end{equation}
which is a characteristic of the WZW action. 
Unlike the case of a sigma model without the Wess-Zumino term, 
parity alone $(z \leftrightarrow \bar{z} $) is not a symmetry. 
In the SIT context, this is due to the slowly varying envelop 
approximation which breaks the 
parity invariance of the Maxwell-Bloch equation. The chiral symmetry generates 
a new solution from a known one. For example, the chiral transform 
of the 1-soliton in Eq.(20) in the resonant case $(\xi = 0)$ 
is again a 1-soliton but with the replacement of constants $a, b$ by  
\begin{equation}
a \rightarrow -{a \over a^{2} + b^{2}} \ , \ 
b \rightarrow { b \over a^{2} + b^{2}} ,
\end{equation}
which changes the shape of the pulse as well as the velocity by
 $v \rightarrow c - v$. 
The currents and the charge also change into
\begin{equation}
J \rightarrow -\bar{J} \ , \ \bar{J} \rightarrow - J \ , \ Q \rightarrow -Q .
\end{equation}
The second case is {\it the dual symmetry} of the 
Krammers-Wannier type\cite{shin1} under
\begin{equation}
\beta \leftrightarrow -\beta \ , \ 
g \leftrightarrow  i\sigma_{1}g \ .  
\end{equation}
Changing the sign of $\beta $ makes the potential upside down so that 
vacuua become maxima of the potential and vice versa. This 
allows us to find a localized 
solution which approaches to a maximum of the potential asymptotically,  
i.e. it reaches to the completely population inverted state. 
In analogy with the nonlinear Schr\"{o}dinger case, we call it a ``dark 
soliton". In practice, the dark soliton for positive $\beta $ 
can be obtained by 
replacing $\beta \rightarrow -\beta \ , \ \bar{z} \rightarrow -\bar{z} $ 
in the (bright) soliton 
solution of the negative $\beta $ case. For example, the dark 1-soliton can 
be written by 
\begin{eqnarray} 
\cos{\varphi }e^{2i\eta } &=& - {b \over \sqrt{(a-\xi)^{2} + b^{2}}}\mbox{tanh} 
(2bz + 2bC\bar{z} ) - i{a-\xi \over \sqrt{(a-\xi )^{2} + b^{2}}} \nonumber \\
\theta &=& -2(a- \xi) (z- C\bar{z} ) -2\xi z
\end{eqnarray} 

Finally, we show that inhomogeneous broadening can be incorporated 
into our formulation naturally with minor modifications. 
We maintain the constraint equation (12) only and modify the zero curvature 
eqution by
\begin{equation}
\left[ \partial + g^{-1}\partial g + \xi g^{-1}Tg - \xi T + 
\tilde{\lambda }T \ , \ \bar{\partial} + 
\left< {g^{-1}\bar{T} g \over \tilde{\lambda } - \xi } \right> \right] = 0
\end{equation}
where the constant $\tilde{\lambda }$ is a spectral parameter 
which becomes $\lambda + \xi $ 
in the sharp line limit. 
The angular bracket 
signifies an average over the spectrum $f(\xi  )$ as given by
\begin{equation}
< \cdots > = \int^{\infty }_{- \infty } ( \cdots )f(\xi  )d\xi .
\label{inhomog}
\end{equation}
 We make the same identification as in Eq.(16) and 
require that $g^{-1}\partial g + \xi g^{-1}Tg - \xi T$ is  
$\xi $-independent since 
$E$ is a $\xi $-independent macroscopic quantity. This results in 
the SIT equation with inhomogenous broadening as given in Eq.(1).
Once again,  by using the dressing method, the 1-soliton can be obtained 
which is the same as in Eq.(20) but with the replacement
\begin{equation}
C = \left< { 1\over (a-\xi )^{2} + b^{2}} \right> \ ,
 \ D = (a-\xi )\left< { 1\over (a-\xi )^{2} + b^{2}} \right> 
- \left< {  a-\xi \over (a-\xi )^{2} + b^{2}} \right> .
\end{equation}
Eq.(14) shows that each frequency $\xi $ corresponds to a specific gauge choice 
in our formulation therefore inhomogenous broadening is equivalent to 
averaging over different gauge fixings. 
This implies that the inhomogenously broadened case can not be treated by a 
single field theory. Nevertheless it is remarkable that the group 
theoretical parametrization of $E, P$ and $ D$ is still valid. 
Another important feature of inhomogenous broadening is that it introduces 
an anomaly term $M$ in the U(1)-current conservation such that 
$\partial \bar{J} + \bar{\partial} J = M$ and
\begin{eqnarray} 
M &=& 2\cot{\varphi } [ \ \cos (\theta - 2\eta ) 
< \sin (\theta - 2\eta ) \sin {2\varphi } > - 
\sin (\theta - 2\eta ) < \cos (\theta - 2\eta ) \sin {2\varphi } >  
\nonumber \\
&& -(\cot^{2}{\varphi }\bar{\partial} \eta  + {1\over 2}\bar{\partial} 
\theta )\partial \varphi   \ ] .
\end{eqnarray} 
This anomaly vanishes in the sharp line limit due to the constraint Eq.(8). 
It also vanishes in the case of 1-soliton and the charge remains conserved. 
This may be compared with the conserved area of topological solitons in the 
presence of inhomogenous broadening. It is an open question whether there 
exists a similar theorem to the area theorem concerning about the stability 
of pulses with phase variation in terms of charge and anomaly. 

In conlusion, we have shown that the SIT equation in the sharp line limit 
can be identified with the complex sine-Gordon equation. This allowed us to 
present various mathematical properties of solutions as well as the 
equation of motion itself. Also, we have shown that the effective field theoretic 
description can be extended to inhomogenously broadened case by averaging 
over different gauge fixings. One of the advantage of the present reformulation of 
the SIT equation is that it clarifies the analytic nature of SIT in terms of 
group theroetical treatment and allows a systematic generalization to other 
multilevel cases than SIT. Also, the concept of U(1)-charge, which is a new 
feature of SIT coming from our for reformulation, can be used in explaining 
earlier numerical result of slow reshaping of pulses in the off-resonant 
case\cite{diels}. Details will appear in a longer version of this Letter\cite{shin3}. 
\vglue .3in 
{\bf ACKNOWLEDGEMENT}
\vglue .2in
This work was supported in part by the program of Basic Science Research, 
Ministry of Education BSRI-96-2442, and by Korea Science and Engineering 
Foundation through CTP/SNU.
\vglue .2in

\end{document}